\documentclass[a4paper]{spie}

\usepackage{graphicx}
\usepackage{psfrag}
\usepackage{amssymb}

\title{Optical Physics of Imaging and Interferometric Phased
Arrays}

\author{Stafford Withington, George Saklatvala, and Michael P.
Hobson \skiplinehalf Cavendish Laboratory, University of
Cambridge, \\ JJ Thompson Avenue, Cambridge, CB3 0HE, UK}

\authorinfo{Further author information: (Send correspondence to S.W.)\\S.W.:
E-mail: stafford@mrao.cam.ac.uk, Telephone: +44 (0)1223 337393\\
G.S.: E-mail: gs262@mrao.cam.ac.uk, Telephone: +44 (0)1223 337247\\
M.H.: E-mail: mph@mrao.cam.ac.uk, Telephone: +44 (0)1223 339992}

\begin{document}

\maketitle

\begin{abstract}Microwave, submillimetre-wave, and
far-infrared phased arrays are of considerable importance for
astronomy. We consider the behaviour imaging phased arrays and
interferometric phased arrays from a functional perspective. It is
shown that the average powers, field correlations, power
fluctuations, and correlations between power fluctuations at the
output ports of an imaging or interferometric phased array can be
found once the synthesised reception patterns are known. The
reception patterns do not have to be orthogonal or even linearly
independent. It is shown that the operation of phased arrays is
intimately related to the mathematical theory of frames, and that
the theory of frames can be used to determine the degree to which
any class of intensity or field distribution can be reconstructed
unambiguously from the complex amplitudes of the travelling waves
at the output ports. The theory can be used to set up a likelihood
function that can, through Fisher information, be used to
determine the degree to which a phased array can be used to
recover the parameters of a parameterised source. For example, it
would be possible to explore the way in which a system, perhaps
interferometric, might observe two widely separated regions of the
sky simultaneously.
\end{abstract}


\section{Introduction}

There is considerable interest in developing phased arrays for
radio astronomy. Projects include the Square Kilometer Array
(SKA), the Low Frequency Array (LOFAR), the Electronic Multibeam
Radio Astronomy Concept (EMBRACE), and the Karoo Array Telescope
(KAT) \cite{A,B,C}. All of these projects are aimed constructing
phased arrays for microwave astronomy, but as technological
capability improves, phased arrays will eventually be constructed
for submillimetre-wave and far-infrared astronomy \cite{D,E}.

Two types of phased array are of interest: (i) imaging phased
arrays, where an array of coherent receivers is connected to a
beam-forming network such that synthesised beams can be created
and swept across the sky; (ii) interferometric phased arrays,
where the individual antennas of an aperture synthesis
interferometer are equipped with phased arrays such that fringes
are formed within the synthesised beams. In this way it is
possible to extend the field of view, to observe completely
different regions of the sky simultaneously, to steer the field of
view electronically, and to observe spatial frequencies that are
not available to an interferometer because the baselines cannot be
made smaller than the diameters of the individual antennas.

It is important to recognise that the synthesised beams of a
phased array need not be orthogonal, and may even be linearly
dependent. Non-orthogonality may be built into a system
intentionally as a way of increasing the fidelity with which an
image can be reconstructed, or it may arise inadvertently as a
consequence of RF coupling and post-processing cross-talk. In some
situations, say in the case of interacting planar antennas, it may
not even be clear how to distinguish one basis antenna from
another, even before the beam-forming network has been connected.

In this paper, we show that the only information that is needed to
determine the average powers, the correlations between the complex
travelling wave amplitudes, the fluctuations in power, and the
correlations between the fluctuations in power at the output ports
of a phased array, or between the output ports of phased arrays on
different antennas, is the synthesised beams. It is not necessary
to know anything about the internal construction of the arrays or
the beam forming networks. Beam patterns may be taken from
electromagnetic simulations or experimental data. In the case of
interferometric phased arrays, the arrays on the individual
antennas do not have to be the same.

The ability to assess the behaviour of a system simply from the
synthesised beam patterns separates the process of choosing the
best beams for a given application from the process of
understanding how to realise the beams in practice. It also
suggests important techniques for simulating phased arrays, and
for analysing experimental data.

\section{Basic Principles}

In practice, an imaging phased array comprises a sequence of
optical components, an array of single-mode antennas, and an
electrical beam-forming network such that each output port
corresponds to a synthesised reception pattern on the input
reference surface, usually the sky. In some cases, the synthesised
reception patterns may be static and designed to give optimum
sampling on a given class of object, whereas in other cases, the
beam-forming network may be controlled electrically to generate a
set of synthesised beams that can be swept across the field of
view. In the case of microwave astronomy, the optical system would
be a telescope, the single-mode antennas would be the horns or
planar antennas of an array of HEMT amplifiers or SIS mixers, and
the beam-forming network would be a system of microwave or digital
electronics.

\begin{figure}
\begin{center}
\includegraphics{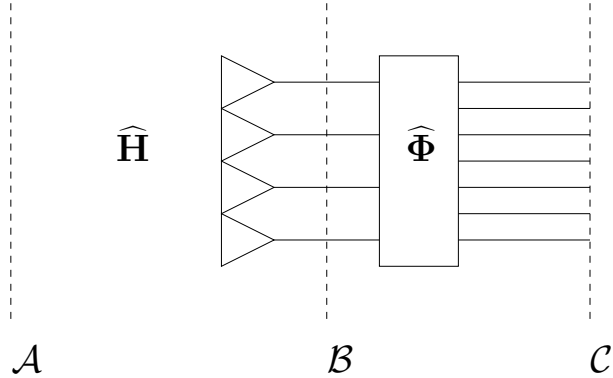}
\end{center}
\caption{A generic phased array having $M$ horns and $P$ ports.}
\end{figure}

Our analysis is based on the generic system shown in Fig.~1.
${\cal A}$ denotes the input reference surface, ${\cal B}$ the
output ports of the horns, and ${\cal C}$ the output ports of the
beam-forming network. We shall assume that an array of $M$ horns
is connected to a beam-forming network having $P$ output ports.
Each of the $P$ ports is thus associated with a reception pattern
on the input reference surface. For simplicity, we shall assume
paraxial optics throughput. When a pseudomonochromatic field,
${\bf x}({\bf r})$, is incident on the system, a set of travelling
waves will appear at ${\cal B}$: we shall denote their complex
amplitudes by $\{ y_{m}: m\in 1,\cdots,M\}$. We shall use the
notion of complex analytic signals throughout, which for most
practical purposes means that one can integrate the final result
over some bandwidth to calculate general behaviour. Likewise, a
set of travelling waves will appear at ${\cal C}$: we shall denote
their complex amplitudes by $\{ z_{p}: p\in 1,\cdots,P\}$. When
$M$ and $P$ are finite, the complex amplitudes can be assembled
into column vectors ${\bf y} \in {\mathbb C}^{M}$ and ${\bf z} \in
{\mathbb C}^{P}$, respectively.

In what follows, it will sometimes be beneficial to represent the
primary variables as abstract vectors. Because the incoming field,
${\bf x}({\bf r})$, is square integrable over the input reference
surface, $\cal A$, it can be represented by a vector $| {\bf x}
\rangle$ in Hilbert space ${\mathbb H}$. The input surface may
extend to infinity, or it may be bounded by an aperture, and
therefore of finite extent. Regions having different shapes and
sizes correspond to different Hilbert spaces. ${\bf y}$ and ${\bf
z}$ can also be represented by abstract vectors, $| {\bf y}
\rangle \in {\ell}^{2}$ and $|{\bf z} \rangle \in {\ell}^{2}$
respectively, where ${\ell}^{2}$ is the space of square-summable
sequences. These definitions lead to two operators, one of which,
$\hat{\bf H}: {\mathbb H} \rightarrow {\ell}^{2}$, maps the
incoming optical field onto the outputs of the horns, and the
other $\hat{\bf \Phi}: {\ell}^{2} \rightarrow {\ell}^{2}$ maps the
outputs of the horns onto the outputs of the beam-forming network.
These individual operators can be combined into a single composite
operator $\hat{\bf T}=\hat{\bf \Phi}\hat{\bf H}:{\mathbb H}
\rightarrow {\ell}^{2}$, which describes the system as a whole.

It can be shown, Appendix A, using only the concepts of {\em inner
product}, {\em operators}, and {\em adjoints} in Hilbert space,
that the complex travelling-wave amplitude appearing at port $p$,
when a field, $ {\bf x} ({\bf r})$, is incident on a system is
given by
\begin{equation}
\label{1_2} z_{p} = \int_{\cal A}{\bf t}_{p}^{\ast} ({\bf r})
\cdot {\bf x} ({\bf r}) \, d^{2} {\bf r} \mbox{,}
\end{equation}
where ${\bf t}_{p} ({\bf r})$ is the functional form of the $p$'th
synthesised reception pattern. $\cal A$ corresponds to the surface
and region over which the Hilbert space is defined. In expressions
such as (\ref{1_2}) we shall show the the complex conjugate
explicitly, even though some notation includes it in the dot
product, as an inner product, implicitly. The reason for the
formality in stating, and indeed deriving (\ref{1_2}), is that
(\ref{1_2}) can be shown to be true even when the beam patterns
are not orthogonal.

The synthesised reception patterns are central to what follows
because, according to (\ref{1_2}), the complex travelling wave
appearing at port $p$ is given by calculating the inner product,
over the input reference surface, between the synthesised
reception pattern ${\bf t}_{p} ({\bf r})$ and the incoming field.
It would be naive to assume, however, that when a system is
illuminated by a field having the form ${\bf t}_{p} ({\bf r})$, a
travelling wave only appears at $p$ . In the case of phased
arrays, the synthesised reception patterns do not have to be
orthogonal, and can even be linearly dependent. Thus, although the
output at a given port is given by the inner product between a
field and a reception pattern, as for orinary antennas, one cannot
assume that there is a one-to-one mapping between the antenna
patterns and the ports.

For example, in the case of Fig.~1, the beam patterns of the
horns, ${\bf h}_{m} ({\bf r})$, are orthogonal, and the outputs of
the horns, $y_{m}$, are given by
\begin{equation}
\label{1_3} y_{m} = \int_{\cal A} {\bf h}^{\ast}_{m} ({\bf r})
\cdot {\bf x}({\bf r}) \, d^{2} {\bf r} \mbox{,}
\end{equation}
but the beam-forming network is described by a linear
operator $\hat{\bf \Phi}$, and therefore
\begin{equation}
\label{1_4}
z_{p} = \sum_{m} \phi_{pm} y_{m}
\mbox{.}
\end{equation}
Substituting (\ref{1_4}) in (\ref{1_3}) we find
\begin{equation}
\label{1_5} z_{p} = \int_{\cal A}  \sum_{m} \phi_{pm} {\bf
h}^{\ast}_{m} ({\bf r}) \cdot {\bf x}({\bf r}) \, d^{2} {\bf r}
\mbox{,}
\end{equation}
which can be cast into the form of (\ref{1_2}) by defining
\begin{equation}
\label{1_6}
{\bf t}_{p}^{\ast} ({\bf r}) = \sum_{m} \phi_{pm}
{\bf h}^{\ast}_{m} ({\bf r})
\mbox{.}
\end{equation}
As expected, the synthesised reception patterns are merely
weighted linear combinations of the horn patterns. The
orthogonality of the synthesised reception patterns can now be
tested through
\begin{equation}
\label{1_7} \int_{\cal A}  {\bf t}_{p}^{\ast} ({\bf r}) \cdot {\bf
t}_{p'} ({\bf r})  \, d^{2} {\bf r} = \sum_{m} \phi_{pm}
\phi_{p'm}^{\ast} \mbox{,}
\end{equation}
where (\ref{1_6}) has been used, together with the orthonormality
of the horn patterns.

In the case where the numbers of horns and ports are
finite, (\ref{1_7}) takes the form of a matrix equation:
\begin{equation}
\label{1_8} \int_{\cal A} {\bf t}_{p}^{\ast} ({\bf r}) \cdot {\bf
t}_{p'} ({\bf r}) \, d^{2} {\bf r} = {\bf \Phi}{\bf
\Phi}^{\dagger} \mbox{.}
\end{equation}
Because ${\bf \Phi}$ is mapping between ${\mathbb C}^{M}$ and
${\mathbb C}^{P}$, ${\bf \Phi}$ is under complete if $P>M$, and
${\bf \Phi}{\bf \Phi}^{\dagger}$ is singular; contrariwise, ${\bf
\Phi}$ is over complete if $P<M$, and ${\bf \Phi}{\bf
\Phi}^{\dagger}$ is not singular. In both cases, except trivially
when certain ports are not connected, the synthesised reception
patterns are not orthogonal, because ${\bf \Phi}{\bf
\Phi}^{\dagger} \neq {\bf I}_{P}$ is not diagonal. In the case
where ${\bf \Phi}$ is unitary, ${\bf \Phi}{\bf \Phi}^{\dagger} =
{\bf I}_{P}$, where ${\bf I}_{P}$ is the identity operator of
dimension $P$, the synthesised reception patterns are orthogonal.
Butler beam forming networks are used in practice to realise this
situation.

In summary, the complex travelling-wave amplitudes appearing at
the output ports of a phased array are found by calculating the
inner products of the incoming field with respect to a set of
synthesised reception patterns, but the synthesised reception
patterns do not have to be orthogonal. Even if a system is
designed to have orthogonal beams, practical issues relating to
coupling and cross talk will cause the beam patterns to be
non-orthogonal at some level. One would, therefore, like to derive
an analysis procedure based on the beam patterns alone, where it
is not necessary to know anything about the internal construction
of the array. For our purposes, we shall assume that the behaviour
of all phased arrays is described by (\ref{1_2}) regardless of
whether it is known how the arrays are constructed or not.

In many cases we are interested in using phased arrays to image
incoherent or partially coherent fields---in the context of
astronomy, although the field on the sky is usually incoherent,
the input reference plane, as far as the phased array is
concerned, may be internal to the optics of the telescope. To this
end, it is convenient to introduce correlation dyadics. We shall
define the correlation dyadic of the incident field according to
\begin{equation}
\label{1_9} \overline{\overline{\bf X}}({\bf r}',{\bf r}) =
\langle {\bf x}({\bf r}) {\bf x}^{\ast}({\bf r}') \rangle \mbox{,}
\end{equation}
where $\langle \, \rangle$ denotes the ensemble average, and ${\bf
x}({\bf r})$ is interpreted as a complex analytic signal. The
tensor $\overline{\overline{\bf X}}({\bf r}',{\bf r})$ contains
complete information about the correlation between the fields at
any two points and in any two polarisations. Once the correlation
dyadic is known, all classical measures of coherence follow.

The correlation between the travelling wave amplitudes at any two
ports can be written $\langle z_{p} z_{p'}^{\ast} \rangle$, or in
matrix form
\begin{equation}
\label{1_10}
{\bf Z} = {\bf z}{\bf z}^{\dagger}
\mbox{,}
\end{equation}
where ${\bf Z} \in {\mathbb C}^{P \times P}$ is a correlation
matrix. The matrix elements of ${\bf Z}$ can be found by using
(\ref{1_2}):
\begin{equation}
\label{1_11} Z_{pp'} = \int_{\cal A} \int_{\cal A} {\bf
t}_{p}^{\ast} ({\bf r}) \cdot \overline{\overline{\bf X}}({\bf
r}',{\bf r}) \cdot {\bf t}_{p'} ({\bf r}')  \, d^{2} {\bf r} \,
d^{2} {\bf r}' \mbox{.}
\end{equation}

Now illuminate the system with an unpolarised, spatially fully
incoherent source
\begin{equation}
\label{1_12}
\overline{\overline{\bf X}}({\bf r}',{\bf r}) =
\overline{\overline{\bf I}} \delta({\bf r}-{\bf r}')
\mbox{,}
\end{equation}
where $\overline{\overline{\bf I}}$ is the dyadic identity operator.
Substituting (\ref{1_12}) in (\ref{1_11}), we find
\begin{equation}
\label{1_13} Z_{pp'} = \int_{\cal A} {\bf t}_{p}^{\ast} ({\bf r})
\cdot {\bf t}_{p'} ({\bf r}) \, d^{2} {\bf r} \mbox{,}
\end{equation}
which shows that, because the synthesised reception patterns are
generally not orthogonal, the travelling waves at the output ports
are correlated. Ultimately, it is these correlations that prevent
one from extracting more and more information from a source, using
a finite number of horns, by synthesising more and more beams.

\section{Frames}

In what follows, we shall need to make use of the mathematical
theory of frames. Suppose for the moment we have some general
monochromatic field $|{\bf x} \rangle$, and that we determine the
inner products with respect to a set of basis vectors  ${\mathbb
T} = \{|{\bf t}_{p} \rangle, \, p \in 1, \cdots, P \}$: $z_{p} =
\langle {\bf t}_{p} | {\bf x} \rangle$. $P$ can extend to
infinity, and we do not make any assumptions about the
orthonormality or linear independence of ${\mathbb T}$. Under what
circumstances can the original vector $|{\bf x}\rangle$, which
represents a continuous function, be recovered unambiguously from
a discrete set of complex coefficients, possibly countable, and
how can this be achieved? In the context of phased arrays, we are
essentially asking under what circumstances can the form of an
incident field be recovered from the outputs of an array, when the
synthesised beams are possibly non-orthogonal and linearly
dependent.

Evaluate the square moduli of the inner products between ${\mathbb
T}$ and any general vector,$|{\bf x}\rangle$, and sum the results.
If there are two constants $A$ and $B$ such that $0<A<\infty$ and
$0<B<\infty$, and
\begin{equation}
\label{2_1} A \parallel {\bf x} \parallel^{2} \, \leq \,
\parallel \hat{\bf T} | {\bf x} \rangle
\parallel^{2} \, \leq B \, \parallel {\bf x}
\parallel^{2}
\mbox{,}
\end{equation}
which can also be written
\begin{equation}
\label{2_2} A \parallel {\bf x} \parallel^{2} \, \leq \, \sum_{p}
\left| \langle{\bf t}_{p} | {\bf x} \rangle_{\mathbb H}
\right|^{2} \, \leq \, B \parallel {\bf x} \parallel^{2} \mbox{,}
\end{equation}
$\forall \, |{\bf x}\rangle \in {\mathbb H}$, then the basis set
${\mathbb T}$ is called a frame with respect to ${\mathbb H}$.
Notice the use of strict inequalities in the allowable values of
$A$ and $B$. In the case where $A \approx B$, the frame is called
a `tight frame' because the inner products for all $|{\bf
x}\rangle \in {\mathbb H}$ lie within some small range, and the
dynamic range needed for inversion is small. When the original
basis is orthonormal, the frame bounds, $A$ and $B$, are equal, as
can be appreciated by inserting $|{\bf x}\rangle = |{\bf
t}_{p'}\rangle$ in (\ref{2_2}). If the frame is over complete, but
normalised, $A$ is a measure of the redundancy in the frame.

If a basis set constitutes a frame, then it can be shown, through
(\ref{2_1}) alone, that $\hat{\bf T}$ is injective, one-to-one,
but not surjective, onto: $\hat{\bf T}$ maps ${\mathbb H}$ onto a
subspace of ${\ell}^{2}$, or when $P$ is finite, a subspace of
${\mathbb C}^{P}$. Consequently, $\hat{\bf T}$ has a left inverse,
$\hat{\bf T}^{-1}$, such that $|{\bf x}\rangle = \hat{\bf
T}^{-1}\hat{\bf T}|{\bf x}\rangle: \forall {\bf x} \in {\mathbb
H}$. $\hat{\bf T}^{-1}$ maps the image of $\hat{\bf T}$,
$\mbox{Im}[{\hat{\bf T}}]$, back onto ${\mathbb H}$, and maps the
null complement of $\mbox{Im}[{\hat{\bf T}}]$,
$\mbox{Im}[{\hat{\bf T}}]^{\bot}$, onto the zero vector in
${\mathbb H}$. The inverse operator is given by
\begin{equation}
\label{2_3}
\hat{\bf T}^{-1} = \left(\hat{\bf T}^{\dagger}\hat{\bf T}\right)^{-1}
\hat{\bf T}^{\dagger} = \hat{\bf S}^{-1}\hat{\bf T}^{\dagger}
\mbox{:}
\end{equation}
$\hat{\bf S}$ is bijective such that $\hat{\bf S}^{-1}$ exists.

It can also be shown that $\hat{\bf T}^{-1}$ satisfies the frame
condition
\begin{equation}
\label{2_4} \frac{1}{B} \parallel {\bf z} \parallel^{2} \, \leq \,
\parallel \hat{\bf T}^{-1} | {\bf z} \rangle \parallel^{2} \, \leq
\, \frac{1}{A} \parallel {\bf z} \parallel^{2} \mbox{.}
\end{equation}
Thus, the more tightly bound the frame, the more tightly bound the
inverse, and the more stable the reconstruction process.

The reconstruction of the original field through $\hat{\bf
T}^{-1}$ can be best implemented by the introduction of dual
vectors. The dual vectors $| \widetilde{\bf t}_{p} \rangle$ of any
given frame ${\mathbb T}$, with respect to Hilbert space ${\mathbb
H}$, can be derived through
\begin{equation}
\label{2_4}
| \widetilde{\bf t}_{p} \rangle = \hat{\bf S}^{-1} |{\bf t}_{p} \rangle
\mbox{.}
\end{equation}
The dual basis set, which we shall call $\widetilde{\mathbb T} =
\{|\widetilde{\bf t}_{p} \rangle, \, p \in 1, \cdots, P \}$, has
the same degree of completeness as the original frame, ${\mathbb
T}$, and therefore it too constitutes a frame with respect to
${\mathbb H}$. Indeed, two representations of any general $|{\bf
x}\rangle$ are possible:
\begin{eqnarray}
\label{2_5} |{\bf x}\rangle & = & \sum_{p} \langle {\bf t}_{p} |
{\bf x} \rangle \, |\widetilde{\bf t}_{p} \rangle \\ \nonumber
|{\bf x}\rangle & = & \sum_{p} \langle \widetilde{\bf t}_{p} |
{\bf x} \rangle \, |{\bf t}_{p} \rangle
\mbox{.}
\end{eqnarray}
If one calculates a set of coefficients by taking the inner
products with a frame, then one inverts the process by
reconstructing the field using the dual vectors. Alternatively, if
one calculates the inner products with respect to the dual
vectors, then one inverts the process by reconstructing the field
using the frame. In the case where the basis vectors are perfectly
complete with respect to ${\mathbb H}$, but not necessarily
orthogonal, the basis is called a Riesz basis, and the basis set
${\mathbb T}$ and dual set $\widetilde{\mathbb T}$ are then
biorthogonal: $\langle \widetilde{\bf t}_{p} | {\bf t}_{p'}
\rangle = \delta_{pp'}: \forall p,p' \in 1,\cdots,P$.

In the case where the basis vectors do not constitute a frame,
that is to say they are under complete with respect to ${\mathbb
H}$, then one can go through the same procedure as before, but
now, when an attempt is made to reconstruct the original field
vector, by using the duals,
\begin{equation}
\label{2_6} |{\bf x}'\rangle = \sum_{p} \langle {\bf t}_{p} | {\bf
x} \rangle \, |\widetilde{\bf t}_{p} \rangle \mbox{,}
\end{equation}
the reconstructed vector $|{\bf x}'\rangle$ cannot, for all
vectors in ${\mathbb H}$, be the same as the original vector
$|{\bf x}\rangle$. It can be shown, straightforwardly, that the
error vector $|{\bf x}\rangle-|{\bf x}'\rangle$ is orthogonal to
the basis vectors; in other words, the solution is as close as
possible within the degrees of freedom available. $|{\bf
x}'\rangle$ is the orthogonal projection of $|{\bf x}\rangle$ onto
${\mathbb S}$, the subspace spanned by the under complete set of
basis vectors. In the context of functions, the reconstructed
field is a least-square fit to the original field. The same
conclusion is reached if the inner products are taken with the
dual vectors of an incomplete basis, and the field reconstructed
using the original vectors.

The relevance to phased arrays is clear, one can measure the
complex outputs of a phased array, and if the reception patterns
constitute a frame with respect to the Hilbert space defined by
the shape, extent, and illumination of the input reference
surface, then the continuous incoming field can be reconstructed
completely from the discrete set of outputs. If the reception
patterns do not constitute a frame, reconstruction leads to the
least square fit that is consistent with the degrees of freedom to
which the phased array is sensitive. If the field is a spatially
fully incoherent source, the number of degrees of freedom in the
field is infinite, even if the field only extends over a finite
region, and an infinite number of horns is needed to realise a
frame. In reality, however, all optical fields only contain a
finite number of degrees of freedom, and therefore frames are, at
least in principle, possible.

\section{Matrix representations}

The theory of frames is intimately related to the operation of
phased arrays. Suppose, for example, that we wish to describe the
behaviour of a phased array by means of a scattering matrix that
relates, for any incoming field, the $P$ reception-pattern
coupling coefficients to the $P$ output ports. One such
representation is simply the $P \times P$ identity matrix, ${\bf
I}_{P}$, because the outputs are given by the coupling
coefficients in any case, but such a representation does not
correctly represent the throughput of the system if the number of
horns is less than the number of ports, because there are fewer
degrees of freedom in the calculated coefficients than the number
of coefficients. The identity matrix does not contain any
information about the physics of the array, which becomes apparent
when one comes to consider internally generated noise.

A better approach is as follows. Using the concept of frames, we
can generate a set of coupling coefficients by representing the
field in terms of the duals of the reception patterns. From
(\ref{1_2})
\begin{equation}
\label{3_1} z_{p} = \int_{\cal A} {\bf t}_{p}^{\ast} ({\bf r})
\cdot \sum_{p'} {z}'_{p'} \, \widetilde{\bf t}_{p'}({\bf r}) \,
d^{2}{\bf r} = \sum_{p'} R_{pp'} {z}'_{p'} \mbox{.}
\end{equation}
where
\begin{equation} \label{3_2} {z}'_{p'} = \int_{\cal A} {\bf
t}_{p}^{\ast} ({\bf r}) \cdot {\bf x}({\bf r}) \, d^{2}{\bf r}
\mbox{.}
\end{equation}
Alternatively, according to (\ref{2_5}), we may represent the
field in terms of the reception patterns themselves:
\begin{equation}
\label{3_3} z_{p} = \int_{\cal A}  {\bf t}_{p}^{\ast} ({\bf r})
\cdot \sum_{p'} \widetilde{z}'_{p'} \, {\bf t}_{p'}({\bf r}) \,
d^{2}{\bf r} = \sum_{p'} R'_{pp'} \widetilde{z}'_{p'} \mbox{.}
\end{equation}
where
\begin{equation}
\label{3_4} \widetilde{z}'_{p'} = \int_{\cal A} \widetilde{\bf
t}_{p'}^{\ast} ({\bf r}) \cdot {\bf x}({\bf r}) \, d^{2}{\bf r}
\mbox{.}
\end{equation}
In (\ref{3_1}) and (\ref{3_3}), $R_{pp'}$ and $R'_{pp'}$ are both
$P \times P$ scattering matrices, which are equally good at
describing the behaviour of the array. Unlike the identity matrix,
however, they can only transmit the same number of degrees of
freedom as the array itself regardless of whether the beam
patterns constitute a frame with respect to the incoming field or
not. They also lead to the appropriate correlations for internally
generated noise, as will be shown later.

Now consider the situation were an optical system is placed in
front of the phased array described by (\ref{3_3}). We wish to
describe the behaviour of the optical system itself in terms of a
scattering matrix. Moreover, we wish to use the synthesised
reception patterns as the basis set on the output side of the
optical system, which we shall now call ${\mathbb T}_{2} = \{|{\bf
t}_{2,p2} \rangle, \, p2 \in \{1, \cdots, P2 \}$, and some other
basis set on the input side of the optical system, which we shall
call ${\mathbb T}_{1} = \{|{\bf t}_{1,p1} \rangle, \, p1 \in \{1,
\cdots, P1 \}$. ${\mathbb T}_{2}$ does not have to be a frame with
respect to all possible field distributions that can appear at the
output, say ${\mathbb H}_{2}$, because we are only interested in
those fields to which the array can couple. ${\mathbb T}_{1}$
does, however, have to be a frame with respect to all possible
field distributions that can appear at the input, because we are
not sure how incoming fields will scatter. If we choose to use the
synthesised reception patterns as the basis for the input
reference surface, we must supplement the set with the complement
of ${\mathbb T}_{2}$ relative to ${\mathbb H}_{1}$. Indeed,
through this process we can define a virtual array whose beams are
${\mathbb T}_{2} \cap [{\mathbb T}_{2}]^{\bot}$. We shall not
develop this idea here.

The behaviour of the optical system can be described by
\begin{equation}
\label{3_5} {\bf x}_{2}({\bf r}_{2}) = \int_{S_{1}}
\overline{\overline{\bf N}}({\bf r}_{2},{\bf r}_{1}) \cdot {\bf
x}({\bf r}_{1}) \, d^{2} {\bf r}_{1} \mbox{,}
\end{equation}
where, ${\bf x}_{2}({\bf r}_{2})$ and ${\bf x}_{1}({\bf r}_{1})$
are the fields on the input and output sides, respectively, and
again, paraxial optics is assumed. Now we can use the dual frame
of  ${\mathbb T}_{1}$, $\widetilde{\mathbb T}_{1}$ say, to
generate a set of expansion coefficients on the input side:
\begin{equation}
\label{3_6} \widetilde{a}_{p1} = \int_{S_{1}} \widetilde{\bf
t}_{1,p1}^{\ast} ({\bf r}_{1}) \cdot {\bf x}_{1}({\bf r}_{1})
 \, d^{2} {\bf r}_{1} \mbox{.}
\end{equation}
and then (\ref{3_5}) becomes
\begin{equation}
\label{3_7} {\bf x}_{2}({\bf r}_{2}) = \int_{S_{1}}
\overline{\overline{\bf N}}({\bf r}_{2},{\bf r}_{1}) \cdot
\sum_{p1} \widetilde{a}_{p1} \, {\bf t}_{1,p1} ({\bf r}_{1}) \,
d^{2} {\bf r}_{1} \mbox{.}
\end{equation}
We can also express the output field in terms of a set of
coefficients
\begin{equation}
\label{3_8} \widetilde{b}_{p2} = \int_{S_{2}} \widetilde{\bf
t}_{2,p2}^{\ast} ({\bf r}_{2}) \cdot {\bf x}_{2}({\bf r}_{2}) \,
d^{2} {\bf r}_{2} \mbox{.}
\end{equation}
Substituting (\ref{3_7}) in (\ref{3_8}) we find
\begin{equation}
\label{3_9} \widetilde{b}_{p2} = \sum_{p1} M_{p2 p1} \,
\widetilde{a}_{p1} \mbox{,}
\end{equation}
where the matrix elements are given by
\begin{equation}
\label{3_10}
 M_{p2p1} = \int_{S_{1}}
\int_{S_{2}} \widetilde{\bf t}_{2,p2}^{\ast} ({\bf r}_{2}) \cdot
\overline{\overline{\bf N}}({\bf r}_{2},{\bf r}_{1}) \cdot {\bf
t}_{1,p1} ({\bf r}_{1}) \, d^{2} {\bf r}_{1} \,  d^{2} {\bf
r}_{2}\mbox{.}
\end{equation}
(\ref{3_10}) is an operator, which is a matrix for finite
dimensional spaces, that maps the field coefficients on the input
side onto the field coefficients on the output side. The operator
describes the process of reconstructing the field in the space
domain, scattering in the space domain, and then projecting the
scattered field onto the output basis set.

If we assume finite dimensionality for all surfaces, and that the
output frame of one optical component is used as the input frame
of the next optical component, then we can cascade a number of
components, $I$, according to
\begin{equation}
\label{3_11}
{\bf M} = \prod_{I} {\bf M}^{i}
\mbox{,}
\end{equation}
where the $\{{\bf M}^{i}: i \in 1\cdots,I\}$ are the scattering
matrices of the individual components. The last component, ${\bf
M}^{I}$, could be the phased array itself, ${\bf R}'$ in
(\ref{3_3}), giving a description of the system as a whole.

Earlier, we showed that it is possible to describe the behaviour
of a phased array in terms of the duals of the reception patterns,
rather than the reception patterns themselves. Equally, we can use
either frames or dual frames on the input and output reference
surfaces of an optical component to generate a variety of
scattering matrices, each of which describes the behaviour of the
component equally well. Moreover, we can choose whether to use
frames or dual frames, or a mixture, in the definition of the
correlation dyadics, thereby generating a variety of equally good
ways of describing correlations \cite{J}. When representing the
process of scattering a partially coherent field through an
optical component, the correlation dyadics should be chosen to
match the bases used for the scattering matrices themselves.

\section{Imaging phased arrays}

\subsection{Imaging field distributions}

It is not possible to construct a phased array that forms a frame
with respect to any undefined complex function, even over a
finite-sized region, because an infinite number of individual
horns would be needed. In reality, however, optical fields have
finite dimensionality, and frames become feasible. Often, a phased
array will be placed on the back of an optical system, and the
role of the phased array is to collect as much of the information
that appears at the output of the optical system as possible. We
now consider whether the outputs of a given phased array form a
frame with respect to any field that can pass through a preceding
optical system.

We have shown previously \cite{K} that the behaviour of paraxial
optical systems is best described using the Hilbert-Schmidt
decomposition of the operator that projects the field at the input
reference surface onto the output reference surface:
$\overline{\overline{\bf N}}({\bf r}_{2},{\bf r}_{1})$ in
(\ref{3_5}). A Hilbert-Schmidt decomposition is needed because
optical systems generally map fields between different Hilbert
spaces, and therefore eigenfunctions are not suitable for
describing behaviour.

Thus, the dyadic Green's function in (\ref{3_5}) becomes
\begin{equation}
\label{4_1} \overline{\overline{\bf N}} ({\bf r}_{2},{\bf r}_{1})
= \sum_{i} \sigma_{i} {\bf u}_{i}({\bf r}_{2}) {\bf
v}_{i}^{\ast}({\bf r}_{1}) \mbox{.}
\end{equation}
After substituting (\ref{4_1}) into (\ref{3_5}) it becomes clear
that the process of scattering a field through an optical system
consists of projecting the incoming field onto the input
eigenfields ${\bf v}_{i}({\bf r}_{1})$, scaling by the singular
values $\sigma_{i}$, and reconstructing the outgoing field through
the outgoing eigenfields $ {\bf u}_{i}({\bf r}_{2})$. It is also
clear, and an intrinsic feature of the Hilbert Schmidt
decomposition, that the field, possibly partially coherent, at the
output reference surface has only a limited number of degrees of
freedom. In the context of (\ref{4_1}), the Hilbert Schmidt
decomposition has only a finite number of singular values that are
significantly different from zero.

What we require is for the synthesised reception patterns of our
phased array to create a frame with respect to the vector space
spanned by the ${\bf u}_{i}({\bf r}_{2})$ having singular values
significantly different from zero: say Hilbert subspace ${\mathbb
S}$. In this case the frame is finite, and could, in principle at
least, be realised by a finite number of horns. How do we
determine whether the synthesised reception patterns constitute a
frame with respect to ${\mathbb S}$?

Suppose that $|{\bf x}_{2}\rangle$ is some general vector in the
Hilbert space ${\mathbb H}_{2}$ at the output reference surface of
an optical system. ${\mathbb S}$ corresponds to that subspace of
${\mathbb H}_{2}$ spanned by the output eigenfields having
singular values greater than some threshold value, say
$\sigma_{min} > \epsilon$. In other words, ${\mathbb S}$ contains,
for all practical purposes, any information that could have been
transmitted through the optical system. The set of output
eigenfields having singular values greater than $\epsilon$ is
$\{|{\bf u}_{i}\rangle :i \in 1,\cdots,I\}$, where $I<\infty$,
because the throughput of the optical system is finite. Now
suppose that we have some other set of vectors $\{|{\bf t}_{p}
\rangle:1,\cdots,P\}$, and wish to determine whether $\{|{\bf
t}_{p}\rangle:1,\cdots,P\}$ constitutes a frame with respect to
${\mathbb S}$. That is to say, if we determine the complex
coupling coefficients between the $|{\bf t}_{p}\rangle$ and any
vector in ${\mathbb S}$, can we recover the vector in ${\mathbb
S}$ without ambiguity?

If $|{\bf x}_{2}\rangle$ is some general vector in ${\mathbb S}$,
then the frame condition reads
\begin{equation}
\label{4_2} A \parallel{\bf x}_{2} \parallel^{2} \leq \sum_{p}
\left| \langle {\bf t}_{p} | {\bf x}_{2} \rangle \right|^{2} \leq
B
\parallel {\bf x}_{2} \parallel^{2} \hspace{5mm} \forall \, |{\bf
x}_{2} \rangle \in {\mathbb S} \mbox{,}
\end{equation}
or, assuming that $|{\bf x}_{2}\rangle$ has been normalised
\begin{equation}
\label{4_3} A \leq \sum_{p} \left| \langle {\bf t}_{p} | {\bf
x}_{2} \rangle \right|^{2} \leq B \hspace{5mm} \forall \, |{\bf
x}_{2}\rangle \in {\mathbb S} \mbox{.}
\end{equation}

For a given set of vectors $|{\bf t}_{p} ({\bf r})\rangle$, the
inner products can be written explicitly, such that (\ref{4_3})
takes the form
\begin{equation}
\label{4_4} A \leq \sum_{p} \left| \int_{\cal A} \, {\bf
t}_{p}^{\ast} ({\bf r}) \cdot {\bf x}_{2} ({\bf r}) \, d^{2} {\bf
r} \right|^{2} \leq B \hspace{5mm} \forall \, {\bf x}_{2} ({\bf
r}) \in {\mathbb S} \mbox{.}
\end{equation}
We can, however, describe ${\bf x}_{2}({\bf r})$ completely in
terms of the output eigenfields
\begin{equation}
\label{4_5} {\bf x}_{2} ({\bf r}) = \sum_{i} a_{i} \, {\bf u}_{i}
({\bf r}) \mbox{.}
\end{equation}

Substituting (\ref{4_5}) into (\ref{4_4}) gives
\begin{equation}
\label{4_6} A \leq  \sum_{p} \left| \sum_{i} E_{pi} a_{i}
\right|^{2} \leq B \hspace{5mm} \forall \, {\bf a} \in {\mathbb
C}^{I} \mbox{,}
\end{equation}
where
\begin{equation}
\label{4_7} E_{pi} = \int_{\cal A}  {\bf t}_{p}^{\ast} ({\bf r})
\cdot {\bf u}_{i} ({\bf r}) \, d^{2} {\bf r} \mbox{.}
\end{equation}
Expanding (\ref{4_6}), we get
\begin{equation}
\label{4_8} A \leq \sum_{ii'} a_{i'}^{\ast} a_{i} R_{i'i} \leq B
\hspace{5mm} \forall \, {\bf a} \in {\mathbb C}^{I} \mbox{,}
\end{equation}
where
\begin{equation}
\label{4_9} R_{i'i} = \sum_{p} E_{i'p}^{\dagger}E_{pi} \mbox{.}
\end{equation}
Or, because the number of basis functions $I$ is finite, ${\bf R}$
can be written as ${\bf R} = {\bf E}^{\dagger}{\bf E}$, where the
elements of ${\bf E}$ correspond to the overlap integrals between
the output eigenfields and the synthesised reception patterns:
(\ref{4_7}). Although, the final relationship expresses a mapping
of a finite dimensional space onto itself, the mapping passes
through a space having infinite dimensions and therefore the
integral in (\ref{4_7}) should be evaluated analytically if at all
possible. The frame condition (\ref{4_8}) then becomes
\begin{equation}
\label{4_10} A \leq {\bf a}^{\dagger} {\bf R} {\bf a} \leq B \, \,
\equiv \, \, A \leq {\bf a}^{\dagger} {\bf E}^{\dagger} {\bf E}
{\bf a} \leq B  \hspace{5mm} \forall \, {\bf a} \in {\mathbb
C}^{I} \mbox{.}
\end{equation}

In order to establish whether $\{{\bf t}_{p}:1,\cdots,P\}$
constitutes a frame with respect to the output eigenfields having
non-zero singular values, we need to determine the limits $A$ and
$B$ by rotating ${\bf a}$ throughout ${\mathbb C}^{I}$. Another
way of thinking about the same problem is that we have some
general ${\bf a}$, and we wish to determine whether it always be
described in terms of the vector space spanned by the set of
vectors ${\bf z}_{p}$, corresponding to the set of all possible
measurements, given the mapping ${\bf E}$.

The operator ${\bf R} = {\bf E}^{\dagger} {\bf E}$ is Hermitian,
and can be diagonalised:
\begin{equation}
\label{4_11} {\bf R} = {\bf W} {\bf \Lambda}{\bf W}^{\dagger}
\mbox{.}
\end{equation}
The frame condition then becomes
\begin{equation}
\label{4_12} A \leq  {\bf a}^{\dagger} {\bf W} {\bf \Lambda}{\bf
W}^{\dagger} {\bf a}  \leq B \hspace{5mm} \forall \, {\bf a} \in
{\mathbb C}^{I} \mbox{.}
\end{equation}
The middle term of (\ref{4_12}) takes on its maximum value when
the vector ${\bf a}$ corresponds to the eigenvector of ${\bf W}$
having the largest eigenvalue: remembering that ${\bf a}$ must
have unit length and therefore can only be rotated. If ${\bf W}$
is degenerate in the largest eigenvalue, there is a range of
vectors that lead to a maximum, but the outcome is still that
$B=\lambda_{max}$. Likewise, (\ref{4_12}) takes on its minimum
value when the vector ${\bf a}$ corresponds the the eigenvector
having the smallest singular value, $A=\lambda_{min}$. If the
smallest singular value is zero, ${\bf R}$ is singular, and
$\{|{\bf t}_{p}\rangle:1,\cdots,P\}$ does not constitute a frame
with respect to ${\mathbb S}$.

The operator ${\bf R} = {\bf E}^{\dagger} {\bf E}$ simply maps the
eigenfield coefficients of the optical system onto the output
ports of the array and then back again onto the eigenfield
coefficients. If the set of basis vectors ${\mathbb T}$ do not
span all possible vectors in ${\mathbb S}$, either because there
are too few of them, or because they do not span the same space,
information is lost when the frame coefficients are calculated,
and the frame is incomplete. It is not possible, therefore, to
recover complete information about the output field of the optical
system from the outputs of the phased array. In this case,
recovering ${\bf a}$ with the dual vectors, and then
reconstructing the field using the eigenfields, will give the best
least square approximation to the field. In reality, because of
the presence of noise a Bayesian method would probably be used to
reconstruct the field.

For infinite-dimensional frames, $P\rightarrow \infty$, we can use
the same procedure, but now we must calculate the eigenvalues of
the matrix ${\bf R}$, where
\begin{equation}
\label{4_14} R_{i'i} = \int_{\cal A} \int_{\cal A}  {\bf
u}_{i'}^{\ast} ({\bf r}) \cdot \overline{\overline{\bf T}} ({\bf
r}',{\bf r}) \cdot {\bf u}_{i} ({\bf r}') \, d^{2}{\bf r} \,
d^{2}{\bf r}' \mbox{,}
\end{equation}
where
\begin{equation}
\label{4_15} \overline{\overline{\bf T}} ({\bf r}',{\bf r}) =
\sum_{p} {\bf t}_{p} ({\bf r}) {\bf t}_{p}^{\ast} ({\bf r}')
\mbox{,}
\end{equation}
and the sum over $p$ extends to infinity. Again, these integrals
should be evaluated analytically. Clearly, in the case where the
frame is complete and orthonormal $ \sum_{p} {\bf t}_{p} ({\bf r})
{\bf t}_{p}^{\ast} ({\bf r}')= \overline{\overline{\bf I}} \delta(
{\bf r}-{\bf r}')$, giving ${\bf R} = {\bf I}$, supporting the
validity of the result.

We now have a measure of how effectively a phased array can image
a complex field; it is easy to show that when a phased array forms
a frame with respect to a fully coherent field at a surface, then
it is also possible to recover completely the spatial correlations
of a partially coherent field at the same surface: essentially
because the natural modes of the partially coherent field lie
within the same Hilbert subspace ${\mathbb S}$.

According to (\ref{2_5}), in the infinite-dimensional case,
\begin{eqnarray}
\label{4_16} z_{p} & = & \int_{\cal A} {\bf t}_{p}^{\ast} ({\bf
r}) \cdot {\bf x}({\bf r}) \, d^{2}{\bf r} \\ \nonumber {\bf
x}({\bf r}) & = & \sum_{p} z_{p} \, \widetilde{\bf t}_{p} ({\bf
r}) \mbox{,}
\end{eqnarray}
which describes the recovery of a coherent field. Incidently,
(\ref{4_16}) also shows that $\sum_{p} \widetilde{\bf t}_{p} ({\bf
r}) {\bf t}_{p}^{\ast} ({\bf r}) = \overline{\overline{I}}
\delta({\bf r})$ for an over complete or perfectly complete frame.
Forming the correlation matrix ${\bf Z}$ and the correlation
dyadic $\overline{\overline{\bf x}}({\bf r}',{\bf r})$, using
(\ref{4_16}), we get
\begin{eqnarray}
\label{4_17} Z_{pp'} & = & \int_{\cal A} \int_{\cal A} {\bf
t}_{p}^{\ast} ({\bf r}) \cdot \overline{\overline{\bf x}}({\bf
r}',{\bf r}) \cdot {\bf t}_{p'} ({\bf r}') \,  d^{2}{\bf r} \,
d^{2}{\bf r}' \\ \nonumber \overline{\overline{\bf x}}({\bf
r}',{\bf r}) & = & \sum_{pp'} Z_{pp'} \, \widetilde{\bf t}_{p}
({\bf r})
 \widetilde{\bf t}_{p'}^{\ast} ({\bf r}')
\mbox{,}
\end{eqnarray}
which describes the recovery of the spatial correlations of a
field from measurements of the cross correlations between the
outputs of a phased array, using the dual beams. (\ref{4_17})
confirms that the correlations of a field can also be recovered,
if the reception patterns constitute a frame.

\subsection{Imaging intensity distributions}

The previous section describes a calculation that can be performed
to find out whether the synthesised reception patterns of a phased
array form a frame with respect to the output eigenfields of an
optical system. This procedure must be used when one is interested
in recovering phase information from the field. Often, however, in
the case of simple imaging, one is only interested in being able
to recover the intensity distribution of a fully incoherent
source. In this case, certain of the beam patterns needed to form
a frame may be created by scanning the array physically across the
source. It seems, however, that different frames are needed
depending on whether one is trying to preserve phase or whether
one is just interested in measuring intensity: we should
distinguish between `field frames' and `intensity frames'.

To this end, assume that the source is fully incoherent and
unpolarised, but that the intensity varies from position to
position. The correlation dyadic of the source then becomes
\begin{equation}
\label{5_1} \overline{\overline{\bf x}}({\bf r}',{\bf r}) =
\overline{\overline{\bf I}} I({\bf r}) \delta({\bf r}-{\bf r}')
\mbox{.,}
\end{equation}
where $I({\bf r})$ is the intensity as a function of position.
Substituting (\ref{5_1}) into (\ref{1_11}) gives
\begin{equation}
\label{5_2} Z_{pp'} = \int_{\cal A}  I({\bf r}) {\bf t}_{p}^{\ast}
({\bf r}) \cdot {\bf t}_{p'} ({\bf r}) \, d^{2} {\bf r} \mbox{.}
\end{equation}
But say that we only measure the diagonal elements of ${\bf Z}$
through the use of total power detectors, then
\begin{equation}
\label{5_3} Z_{pp} = \int_{\cal A}  I({\bf r}) k_{p} ({\bf r}) \,
d^{2} {\bf r} \mbox{,}
\end{equation}
where $k_{p} ({\bf r}) = {\bf t}_{p}^{\ast} ({\bf r}) \cdot {\bf
t}_{p} ({\bf r})$. Thus, for an incoherent source, the output
powers of the individual ports of a phased array are related to
the intensity distribution of the source through a set of inner
products with the functions $\{k_{p} ({\bf r}):p \in
1,\cdots,P\}$.

If the goal is to reconstruct the intensity distribution of a
source, one could ask whether the basis $ \{k_{p} ({\bf r}): p \in
1,\cdots,P\}$ forms a frame with respect to the Hilbert space
defining the range of possible intensity distributions. There is a
problem, however, because in assuming that the source field is
spatially incoherent, we assumed that the intensity is a member of
an infinite dimensional space. To answer the question as to
whether the phased array is suitable for recovering intensity, we
must define more clearly the vector space of intensity
distributions that is of interest.

One possible approach is to describe the intensity distribution as
a weighted linear combination of basis functions, $\psi_{n} ({\bf
r})$. These functions could, for example, be radial basis
functions, wavelets, or delta functions at sample points. If
chosen carefully, these functions need not correspond to a single
region, but could represent a number of spatially separated
regions that one wishes to image simultaneously. If we
characterise the space of intensity distributions according to
\begin{equation}
\label{5_4} I({\bf r}) = \sum_{n} a_{n} \, \psi_{n} ({\bf r})
\mbox{,}
\end{equation}
then the powers recorded at the output of the phased array become
\begin{equation}
\label{5_5} Z_{pp} = \sum_{n} a_{n} \, F_{pn} \mbox{,}
\end{equation}
where
\begin{equation}
\label{5_6} F_{pn} = \int_{\cal A}  k_{p} ({\bf r}) \psi_{n} ({\bf
r}) \, d^{2} {\bf r} \mbox{.}
\end{equation}
The frame condition then reads
\begin{equation}
\label{5_7} A \leq {\bf a}^{T} {\bf F}^{T} {\bf F} {\bf a} \leq B
\hspace{5mm} \forall \, {\bf a} \in {\mathbb C}^{N} \mbox{,}
\end{equation}
where $A$ and $B$, and hence the tightness of the frame, can be
determined by finding the eigenvalues of ${\bf F}^{T}{\bf F}$, or
the SVD of ${\bf F}$. In the case where the basis functions
correspond to sample points ${\bf r}_{n}$, we have $\psi_{n}({\bf
r}) = \delta({\bf r}-{\bf r}_{n})$ and $F_{pn} = k_{p}({\bf
r}_{n})$. Clearly, the original intensity distribution can be
found, to within the degrees of freedom $N$, by using the dual
vectors of $k_{p} ({\bf r})$, namely $ \widetilde{k}_{p} ({\bf
r})$, defined in the space of $\psi_{n}({\bf r})$. Usually,
however, for stochastic sources, and when noise is included, a
Bayesian method would be used to recover images.

It is instructive to see how this form of analysis compares with,
and is applicable to, multimode bolometric imaging arrays
\cite{L}. It has been \cite{M} shown that the expectation value,
$\mbox{E}[P]$,  of the output of essentially any multimode
bolometric detector is given by
\begin{equation} \label{5_8} \mbox{E}[P] = \frac{1}{2 \pi}
\int_{-\infty}^{+\infty} \int_{\cal A} \int_{\cal A}
\overline{\overline{X}}({\bf r},{\bf r}',\omega) \odot
\overline{\overline{T}}({\bf r},{\bf r}',\omega) \, d^2{\bf r} \,
d^{2}{\bf r}' \, d \omega \mbox{,}
\end{equation}
where $\overline{\overline{T}}({\bf r},{\bf r}',\omega)$ is a
tensor that characterises completely the physics of the detector,
and can include any optical system and filters that precede the
detector. $\odot$ denotes the full tensor contraction to a single
real variable, and $\omega$ indicates the frequency dependence of
the tensors. If we now assume a completely unpolarised, incoherent
source, as described by (\ref{5_1}), then the output of the
detector becomes
\begin{equation}
\label{5_9} \mbox{E}[P] = \frac{1}{2 \pi} \int_{-\infty}^{+\infty}
\int_{\cal A} I ({\bf r},\omega) k({\bf r},\omega) \, d^2{\bf r}
\, d \omega \mbox{,}
\end{equation}
where $k({\bf r},\omega)$ is the sum of the diagonal elements of
$\overline{\overline{T}}({\bf r},{\bf r}',\omega)$ evaluated at a
single position. In other words, it gives the output of a
multimode bolometric detector as a function of position.
(\ref{5_9}) has precisely the same form as (\ref{5_3}), and
therefore one can use, as before, the theory of frames to
determine the degree to which an array of multimode bolometric
detectors creates a frame with respect to a given class of
intensity distributions. In fact, a multimode bolometric detector
can be created from a phased array by measuring the power arriving
at each output port, multiplying each of the measurements by some
weighting factor, and then adding all of the results together.
This comparison will become important when we come to consider
interferometric phased arrays, and the fluctuations and
correlations that appear at the outputs of phased arrays.

To finish this section, it is important to stress that the above
analysis applies only when the source is fully spatially
incoherent; it does not apply to recovering the intensity
distribution of a field that is partially coherent; in that case,
the results of the previous section should be used. Because the
source must be fully incoherent, the analysis applies to primary
sources, although the critical point is that the coherence length
must be smaller than the interval over which the reception
patterns change appreciably. Thus the analysis is appropriate for
many practical situations, but is not applicable, for example, in
the case of recovering the intensity in the focal plane of a low
throughput optical system.

\section{Interferometric Phased Arrays}

We now consider the behaviour of phased arrays in the context of
interferometry. By `interferometric phased array' we mean any
interferometer where the individual elements are equipped with
phased arrays for the purpose of creating a number of primary
beams on the sky simultaneously. Central to the analysis is the
observation that an interferometric phased array is essentially a
bolometric interferometer \cite{K}, where the individual phased
arrays are associated with a number of natural modes, which are
equivalent to the natural modes of a multimode bolometer.

Suppose that a number of telescopes configured as an
interferometer, and that each telescope is equipped with a phased
array. We know that each port of the beam forming network is
associated with a synthesised reception pattern on the sky, but
equally, we recognise that, in general, these synthesised beams
are not orthogonal. In this context, we have already described the
mapping ${\bf T}: |{\bf x}\rangle \longmapsto |{\bf z}\rangle$ as
${\bf T}:{\mathbb H} \rightarrow {\ell}^{2}$. The phased array
acts as a linear operator between two Hilbert spaces: one being
the space of square integrable functions over the input reference
surface, and one being the space of square summable complex
sequences. For any real system, this operator must be Hilbert
Schmidt as the amount of information that can be transmitted is
finite. The integral operator can be written
\begin{equation}
\label{6_1} z_{p} = \int_{\cal A} \sum_{i} \sigma_{i} \, {\bf
U}_{i}(p) {\bf V}_{i}^{\ast} ({\bf r}) \cdot {\bf x}({\bf r}) \,
d^{2}{\bf r} \mbox{,}
\end{equation}
which is the equivalent of (\ref{4_1}), allowing for the fact that
the output is a discrete vector.

The operation of a phased array can therefore be regarded as first
mapping the incoming field onto the input eigenfields, ${\bf
V}_{i} ({\bf r})$, which are orthogonal, scaling by the singular
values, $\sigma_{i}$, and then reconstructing the complex
travelling wave amplitudes at the output through the basis vectors
${\bf U}_{i}(p)$, which are also orthogonal. Those input
eigenfields associated with non-zero singular values span the
field distributions at the input to which the phased array is
sensitive, and those output eigenfields associated with non-zero
singular values span the vectors at the output to which the phased
array can couple. Moreover, it can be shown the the input
eigenfields associated with different telescopes are mutually
orthogonal, and therefore the eigenfields of different telescopes
can be combined to form a single large, orthonormal, composite
basis set that can be used to propagate any partially coherent
field through a complete interferometer. The input eigenfields
actually describe those field distributions that can be traced to
the output ports and then back again onto the sky unchanged in
form; they are the eigenfunctions of complete round trips.

The analysis of an interferometric phased array proceeds as
follows. Calculate the Hilbert-Schmidt decomposition of each
telescope, and pick out those eigenfields having non-zero singular
values above the threshold, $\epsilon$, of interest. Place phase
slopes on the eigenfields in accordance with the baselines of the
interferometer. This procedure has already been described in
detail in the context of bolometric interferometers, and will not
be repeated here \cite{N,O}. The elements of the correlation
matrix describing the correlations between the different output
ports of the phased arrays on different telescopes then become
\begin{equation} \label{6_2} Z_{pp'} = \sum_{i} \sum_{i'}
\sigma_{i} \sigma_{i'} {\bf U}_{i}(p) {\bf U}_{i'}^{\ast}(p')
\int_{\cal A} \int_{\cal A} {\bf V}_{i}^{\ast} ({\bf r}) \cdot
\overline{\overline{X}}({\bf r}',{\bf r}) \cdot {\bf V}_{i'} ({\bf
r}') \, d^{2}{\bf r} \, d^{2}{\bf r}' \mbox{,}
\end{equation}
where the sums over the eigenfields $i$ and $i'$ extend to all
telescopes. In the case where the source is spatially incoherent
and unpolarised, (\ref{6_2}) becomes
\begin{equation} \label{6_3} Z_{pp'} = \sum_{i} \sum_{i'}
\sigma_{i} \sigma_{i'} {\bf U}_{i}(p) {\bf U}_{i'}^{\ast}(p')
 \int_{\cal A} I({\bf r}) {\bf V}_{i}^{\ast} ({\bf
r}) \cdot {\bf V}_{i'} ({\bf r}') \, d^{2}{\bf r} \mbox{.}
\end{equation}
${\bf Z}$ contains complete information about the correlations
between the outputs of phased arrays on the same and different
telescopes in terms of the intensity, $I({\bf r})$, of the field
on the sky, and can be used to produce simulated fringes
\cite{N,O}.

Given that the combined set of input eigenfields of all antennas
spans completely the fields on the sky to which an interferometer
is sensitive, it would also be straightforward to determine
whether a given set of baselines and phased arrays comprise a
frame with respect to some class of intensity distributions. In
reality, the Fourier plane is rarely sampled completely, and in
any case the calculation of the frame bounds would be
computationally intensive.

\section{Noise}

We have described a scheme for analysing the behaviour of phased
arrays and interferometric phased arrays, and it would be
desirable to include noise. Ideally, one should be able to model
any internally generated noise by using the synthesised reception
patterns alone. Also, we need to determine not just the noise
power appearing at the outputs of an array, but the fluctuations
and correlations in the fluctuations in the power arriving at the
output ports. After all, it is the fluctuations that determine the
sensitivity with which a measurement can be made.

If the receiver noise temperatures associated with the primary
horns are known, and equal, then the procedure is straightforward.
By definition, the noise temperature of a receiving channel is the
temperature that a matched source would need to have in order to
generate the same output power as a noiseless, but otherwise
identical, system.

Using (\ref{3_1}) in matrix form, the correlations between the
outputs of a phased array are given by
\begin{equation}
\label{7_1} {\bf Z} = {\bf R}\left[{\bf Z}'+{\bf
Z}'_{N}\right]{\bf R}^{\dagger} \mbox{,}
\end{equation}
where ${\bf Z}'_{N}$ is the correlation matrix of an equivalent
set of noise sources at the input, one for each synthesised beam,
which are incoherent with respect to the true source ${\bf Z}'$.
Strictly speaking, (\ref{7_1}) assumes that all of the noise is in
the same spatial modes as the signal, which of course need not be
true. (\ref{7_1}) can be extended easily to account for the more
general case, but we shall not do so here.

To find ${\bf Z}'_{N}$ we simply need to project a uniform
background source having an intensity that is equal to the noise
temperature onto the synthesised beams; using (\ref{3_2}) we get
\begin{equation} \label{7_2}  Z'_{N,pp'} = \int_{\cal A} T_{n} \,
{\bf t}_{p}^{\ast} ({\bf r}) \cdot {\bf t}_{p'} ({\bf r}) \, d^{2}
{\bf r} \mbox{.}
\end{equation}
The diagonal terms of $Z'_{N,pp'}$ give the noise temperatures
that must be associated with each of the synthesised beams, and
importantly, the off-diagonal terms give the correlations between
them. If the beams are orthogonal, the noise sources are
uncorrelated, and one returns to the original definition of noise
temperature.

\section{Correlations and fluctuations}

The net outcome of the previous sections is the ability to
calculate the correlations that appear at the output ports of a
phased array, or the output ports of phased arrays on different
telescopes, from knowledge of the synthesised, possibly
non-orthogonal and linearly dependent, reception patterns. Once
this information is known, many measurable quantities follow;
including average powers, fluctuations in power, field
correlations, and fluctuations in power correlations between
different ports. Moreover, these matrices contain the Hanbury
Brown-Twiss correlations associated with phased arrays. The
expressions that follow have been derived previously
\cite{F,M,P,Q}, but will be reproduced here for completeness.

Rather than simply determining the output powers of individual
ports, it is more general to consider detectors that measure the
powers at a number of ports simultaneously according to some
weighting vector. Characterise each weighted combination of
detectors by diagonal matrix ${\bf W} \in {\mathbb C}^{P \times
P}$, where the diagonal elements are the factors that weight the
sensitivities of the individual detectors that are connected to
the phased array. Under these circumstances, and assuming
radiation whose intrinsic reciprocal coherence time is much
greater than the bandwidth of the system $\Delta \omega$, which is
valid for radio astronomy systems, it can be shown that the
expectation value of the power $\mbox{E}[P]$, recorded by a
detector combination is given by
\begin{equation}
\label{8_1} \mbox{E}[P]  =  \int \mbox{Tr} {\bf Z}{\bf W} \, d
\omega \mbox{.}
\end{equation}
Likewise, assuming Gaussian statistics, the fluctuations in the
output $C_{ss}$ and the correlations between the fluctuations of
two outputs $C_{st}$ are given by
\begin{eqnarray}
\label{8_2} C_{ss} & = & \frac{1}{\tau} \int \mbox{Tr} {\bf Z}{\bf
W}_{s}{\bf Z}{\bf W}_{s} \, d \omega  \\ \nonumber C_{st} & = &
\frac{1}{\tau} \int  \mbox{Tr} {\bf Z}{\bf W}_{s}{\bf Z}{\bf
W}_{t} \, d \omega \mbox{.}
\end{eqnarray}
where all quantities are allowed to be a function of frequency.
The only restriction on (\ref{8_2}) is that the post-detection
integration time $\tau >> 1/\Delta \omega$ where ${\bf Z}(\omega)
\rightarrow 0$ for frequencies outside of $\Delta \omega$, which
is necessary for all astronomical instruments.

These expressions can be extended to describe the quantum
mechanical behaviour of phased arrays, as has been done for
bolometric interferometers \cite{Q}. The bolometric interferometer
model did not, however, include the Poisson limit for low photon
occupancies\cite{T}. In this volume\cite{R}, we describe how one
can add a single term to (\ref{8_2}) to create a statistical
mixture that includes the Poisson noise of photon counting. Once
the additional term is included, it is possible to take into
account the transition from fully Poisson to fully bunched
behaviour, as the photon occupancies of the incoming modes
increase, as one moves from infrared to submillimetre wavelengths.
It is entirely possible, therefore, to modal the
quantum-statistical behaviour of phased arrays at all wavelengths.

Expressions (\ref{8_1}) and (\ref{8_2}) offer a further
possibility of considerable importance. Clearly, we have a
numerical procedure for determining the expectation values and the
covariances of the powers that arrive at the output ports of an
imaging, or interferometric, phased array when a source is
present. A discretised version of the model has already been
published\cite{F}. The model takes into account noise, and can be
extended easily to include quantum effects. Also, the beam
patterns do not have to be orthogonal. (\ref{8_1}) and (\ref{8_2})
therefore make it straightforward to set up a likelihood function
for the outputs that would be recorded when some class of source
is observed. Obviously the likelihood function would contain the
signal, its fluctuations, and any instrumental noise, including
quantum effects, as well as the Hanbury Brown-Twiss correlations
between pixels. The source may be as simple as a single incoherent
Gaussian on the sky, or if one is trying to design a phased array
that can observe two different regions of the sky simultaneously,
it could be two highly separated Gaussians. Any other
parameterised source distribution could be used; for example,
Cauchy functions are often used in astronomy to parameterise
Sunyaev-Zel'dovich emission from clusters of galaxies, and in
(\ref{8_2}), we used a Gauss-Schell source as a convenient way of
parameterising general partially coherent fields.

On the basis of the likelihood functions, one could then derive
numerically, the Fisher information matrix \cite{R,S}, from which
the covariance matrix of the source parameter estimators could be
found. We have already started to apply this technique, in a
completely different context, for understanding the design of
bolometric imaging arrays: see Saklatvala\cite{R}, in this volume.
In order words, one could determine the minimum errors, and the
confidence contours, that could be achieved when determining the
parameters of sources. Exploring how these errors change as the
design of a phased array changes, say by packing more and more
overlapping beams into a finite region, would be of considerable
interest, and the result should be related, in some way, to the
effectiveness with which the array forms a frame with respect to
the incoming field distributions, or intensity distributions, of
interest.

\section{Conclusion}

We have studied the functional behaviour of imaging phased arrays
and interferometric phased arrays, and shown that their operation
is closely related to the mathematical theory of frames. In order
to calculate the behaviour of an imaging phased array, or an
interferometric phased array, it is only necessary to know the
synthesised reception patterns, which may be non-orthogonal and
linearly dependent. It is not necessary to know anything about the
internal construction of the array itself. As a consequence, data
can be taken from experimental measurements or from
electromagnetic simulations. The theory of frames allows one to
assess, in a straightforward manner, whether the outputs of a
phased array contain sufficient information to allow a field or
intensity distribution to be reconstructed in an unambiguous way.

Our model also allows straightforward calculation of quantities
such as the correlations in the fluctuations at the output ports
of phased arrays. The theory of interferometric phased arrays is
almost identical to the theory of multimode bolometric
interferometers, and therefore, recently developed techniques for
modelling bolometric interferometers can be applied to phased
arrays also: including quantum statistics. The work opens up the
important possibility of constructing likelihood functions that
enable the covariance matrices of source-parameter estimators to
be determined. Thus, for example, one could explore the
possibility of enhancing source reconstruction by packing in more
and more overlapping synthesised beams into a region, or widely
separated regions, of finite size. Any enhancement of the accuracy
with which source parameters can be recovered, will be related,
and to some extent determined, by the degree to which the beam
patterns of the array form a frame with respect to all possible
incoming field distributions.

In a later paper, we shall use the concepts described here, and
the numerical techniques reported previously, to simulate and
assess the behaviour of interferometric phased arrays when
different optical systems and beam-forming networks are used.

\begin{appendix}

\section{Deriving an expression for the travelling waves at the
outputs of a phased array}

Suppose that some field, $|{\bf x}\rangle$, is incident on a
phased array, we can represent a measurement of the amplitude and
phase of the travelling wave $z_{p}$ at $p$, relative to a
normalised reference signal $z_{p}'$, by the inner product
$\langle {\bf z}_{p}' | {\bf z} \rangle_{{\ell}^{2}} = z_{p}^{'
\ast} z_{p}$, where $| {\bf z}_{p}'\rangle$ is a vector
corresponding to a measurement at port $p$ alone. For example, the
measurement could be carried out by homodyne mixing the travelling
wave $z_{p}$ with a reference oscillator $z_{p}'$ at $p$, and then
low-frequency filtering the result. Introducing the linear
operator $\hat{\bf T}$, leads to a measurement of $\langle {\bf
z}_{p}' | \hat{\bf T}{\bf x} \rangle_{{\ell}^{2}}$, which by
definition of the adjoint, can be written $\langle \hat{\bf
T}^{\dagger} {\bf z}_{p}' | {\bf x} \rangle_{\mathbb H}$. In other
words the inner product between $|{\bf x} \rangle$ and the field
distribution represented by $| \hat{\bf T}^{\dagger}{\bf z}_{p}'
\rangle$ gives the same result as the measurement, but now the
inner product is evaluated at the input reference surface. We
shall call $| \hat{\bf T}^{\dagger} {\bf z}_{p}' \rangle$ the
synthesised reception pattern of port $p$. The canonical inner
product in ${\mathbb H}$ takes the form
\begin{equation}
\label{A1_1} \langle \hat{\bf T}^{\dagger} {\bf z}_{p}' | {\bf x}
\rangle_{\mathbb H} = \int_{\cal A}  z_{p}^{' \ast} {\bf
t}_{p}^{\ast} ({\bf r}) \cdot {\bf x}({\bf r})  \, d^{2} {\bf r}
\mbox{,}
\end{equation}
where ${\bf t}_{p} ({\bf r})$ is the functional form of the
synthesised reception pattern, because the result must be equal to
$z_{p}^{' \ast} z_{p}$, and therefore conjugate linear in
$z_{p}'$. In (\ref{A1_1}), the integral over $\cal A$ corresponds
to the input reference surface, and extends over the region
associated with Hilbert space ${\mathbb H}$. Finally, because
(\ref{A1_1}) must be equal to $z_{p}^{' \ast} z_{p}$, we have an
expression that relates the complex amplitude of the travelling
wave at $p$ to the incident field:
\begin{equation}
\label{A1_2} z_{p} = \int_{\cal A}{\bf t}_{p}^{\ast} ({\bf r})
\cdot {\bf x} ({\bf r}) \, d^{2} {\bf r} \mbox{,}
\end{equation}
It is clear from (\ref{A1_2}) that the synthesised reception
pattern is the complex conjugate of what would be measured in an
experiment where a point source is swept over the input surface.
The key point is that (\ref{A1_2}) is valid even when the beams
are not orthogonal.

\end{appendix}

\end{document}